\documentclass[reprint,aps,prl]{revtex4-2}

\usepackage{dcolumn}
\usepackage{bm}
\usepackage{amsmath}
\usepackage{amsfonts}
\usepackage{color}
\usepackage[dvipsnames]{xcolor}
\usepackage{graphicx,graphics}
\usepackage{epstopdf}
\usepackage{pifont}
\usepackage{etoolbox}
\usepackage{indentfirst}
\usepackage{braket}
\usepackage{ulem}
\usepackage{dsfont}
\usepackage{slashed}
\usepackage[chorus]{emf}

\usepackage{bookmark,hyperref}
\hypersetup{
	colorlinks,
	citecolor=blue,
	filecolor=blue,
	linkcolor=blue,
	urlcolor=blue
}

\def\bb{\begin{equation}}
\def\ee{\end{equation}}

\def\5{\hspace*{5mm}}

\begin{document}

\title{On the analogue of Einstein--Gauss--Bonnet theory in 3+1 dimensions}

\author{Giorgi Tukhashvili}
\email{giorgit@princeton.edu}
\affiliation{Department of Physics, Princeton University, Princeton, NJ, 08544, USA}


\begin{abstract}

Higher curvature corrections to the Einstein--Hilbert term may play an important role in probing the strong-field regime of gravity. In this letter, we demonstrate that the local effective action reproducing the trace anomaly can resemble the Einstein--Gauss--Bonnet theory in four dimensions on specific backgrounds. The two key observations support this claim: 1) the covariant equation of the trace anomaly coincides with the trace of the metric variation in Einstein--Gauss--Bonnet theory, and 2) on the FRW space-time, the Friedmann-like equations in both frameworks coincide, with this correspondence extending to the quadratic and cubic perturbations. As an intrinsically four-dimensional construct, the trace anomaly effective action emerges as a promising framework for exploring higher curvature corrections to Einstein's General Relativity in a self-consistent manner.

\end{abstract}

\maketitle


\noindent
{\it Introduction.} --
The addition of quadratic curvature corrections to the Einstein--Hilbert (EH) term has significant implications both at a phenomenological level, such as providing a natural mechanism for inflation \cite{Starobinsky:1980te} -- and at a fundamental level -- for instance, rendering gravity renormalizable \cite{Stelle:1976gc}. However, in four dimensions, it is impossible to incorporate such terms into the EH action without introducing additional degrees of freedom.
Lovelock's theorem \cite{Lovelock:1971yv,Lovelock:1972vz} states that the Einstein--Hilbert theory, supplemented with Cosmological Constant, is the unique theory in four-dimensional space-time (4D) that satisfies the following properties: a) is local and diffeomorphism invariant, b) the equations of motion are of the second order in space-time derivatives, c) the gravitational sector propagates only the transverse and traceless helicity-2 modes.
In \cite{Glavan:2019inb,Tomozawa:2011gp,Cognola:2013fva}, an attempt was made to bypass Lovelock's theorem by introducing a divergent coefficient in front of the Gauss--Bonnet invariant that was supposed to ``regularize" its vanishing variation in 4D. However, it was soon realized that such a procedure was inconsistent \cite{Bonifacio:2020vbk,Gurses:2020ofy,Gurses:2020rxb}, as it violated the Bianchi identity and, as a result, the diffeomorphism invariance. 

A classical theory that is invariant under the Weyl re-scaling of the metric, $g_{\mu \nu} \rightarrow e^{2 \sigma} g_{\mu \nu}$, has a vanishing energy-momentum tensor ($\sigma$ is a local parameter). This symmetry is anomalously violated through the quantum loops, and as a consequence, the trace of the energy-momentum tensor is no longer zero in a quantum theory \cite{Capper:1975ig,Duff:1977ay}, known as the trace (Weyl) anomaly.
In the early 80s of the past century, it was shown that it is possible to reproduce the trace anomaly from a local effective action at the expense of an additional scalar degree of freedom \cite{Riegert:1984kt,Fradkin:1983tg}. However, these papers did not address the self-consistency of the scalar, which lacked a kinetic term and was thus strongly coupled. A canonical kinetic term for the new scalar spoils the anomalous trace equation. Recently, it was shown that it is possible to give a kinetic term to the scalar field without altering the structure of the anomaly \cite{Gabadadze:2023quw} (the same action was also obtained in \cite{Fernandes:2021dsb} on different grounds). 

In this letter, we demonstrate that the effective action describing the trace anomaly \cite{Gabadadze:2023quw} can resemble the Einstein--Gauss--Bonnet (EGB) theory in 4D. The first hint to this claim is that equation (\ref{cov_anom_eq}), describing the trace anomaly, coincides with the trace of the metric variation in EGB. This is a background-independent statement. In addition, we show that on the FRW background, there exists a dynamical solution of the above-mentioned scalar, such that the equations of motion for the gravitational field are identical to those obtained in EGB. Also, the lagrangians of both the scalar and tensor perturbations, within the effective action of the trace anomaly on the quadratic and cubic levels, contain the perturbations that exist in EGB. 
It is interesting to check if such solutions of the scalar exist on other backgrounds, but in this letter, we will focus only on the spatially flat FRW.
Lovelock's theorem does not apply to our case, as we are introducing an additional scalar degree of freedom -- the anomalyon \cite{Gabadadze:2024wgj}.
We will be using (+,+,+) conventions from \cite{Misner:1973prb}, in addition we take $M_{Pl}^2 = 1/(8 \pi G) = 2$.


\vspace{.1in}
\noindent
{\it Higher dimensional Einstein--Gauss--Bonnet theory} -- is defined by the following action:
\bb\label{Zwiebach_action}
\mathcal{L}^{EGB} = \sqrt{g} \left[ R + \frac{\kappa^2 ~ E}{(n-3)(n-4)}
 - \frac12 \left( \partial \omega \right)^2 - V (\omega) \right] .
\ee
Here $R$ is the Ricci scalar, $E$ is the Gauss--Bonnet invariant,
\bb
E = R_{\mu \nu \rho \sigma} R^{\mu \nu \rho \sigma} - 4 R_{\mu \nu} R^{\mu \nu} + R^2\,,
\ee 
and the canonical scalar field, $\omega$, is introduced as a simple source.
The lagrangian (\ref{Zwiebach_action}) is defined in any dimension $n > 4$. The inverse power of $(n-3)(n-4)$ is chosen for convenience and has nothing to do with the ``regularization" discussed in \cite{Glavan:2019inb}. The constant $\kappa^2 >0$, as dictated by the Heterotic string effective actions \cite{Zwiebach:1985uq,Gross:1986mw}. The trace of the metric variation of (\ref{Zwiebach_action}):
\bb\label{trace_EGB}
(n-2) R + \frac{\kappa^2}{n-3} E = - \left[ T \right] .
\ee
Here $\left[ T \right]$ is the trace of the canonical scalar field's energy-momentum tensor:
\bb
T_{\mu \nu} \equiv \partial_\mu \omega \partial_\nu \omega -  g_{\mu \nu} \Big( \frac12 \left( \partial \omega \right)^2 + V (\omega) \Big) .
\ee
Let's restrict the space-time background to spatially flat FRW:
\bb\label{FRW_ansatz}
ds^2 = g_{\mu \nu}^{FRW} dx^\mu dx^\nu = a^2 \Big( - d\tau^2 + d \vec{x}^2 \Big) \,.
\ee
The time-time component of the metric variation of (\ref{Zwiebach_action}):
\bb\label{FriedmanN_EGB}
  \frac12 (n-1) (n-2) \Big( \mathcal{H}^2 + \kappa^2 \frac{\mathcal{H}^4}{a^2} \Big) = \frac12 T_{\tau \tau} .
\ee
Here $\mathcal{H} \equiv a'/a$ and $' \equiv \partial/\partial \tau$. Let's assume that there exist solutions for which the metric takes the form (\ref{FRW_ansatz}) and the canonical scalar field $\omega = \omega_0 (\tau)$. The perturbations on these solutions are defined as:
\bb\label{perts_on_FRW}
g_{\mu \nu} = g_{\mu \nu}^{FRW} + a^2 h_{\mu \nu} + a^2 \bar{h}_{\mu \nu} ,
\ee
\bb\label{pert_on_omega}
\omega = \omega_0 + \theta .
\ee
Here, $h_{\mu \nu}$ is the transverse and traceless, helicity-2 part, while $\bar{h}_{\mu \nu}$ encodes the scalar perturbations \cite{Mukhanov:1990me}:
\bb
\bar{h}_{00} = - 2 \phi ,
\5\5\5
\bar{h}_{0 i} = \partial_i \mathcal{B} ,
\ee
\bb
\bar{h}_{i j} = - \left( 2 \psi + \frac{1}{3} \partial_k^2 \emf \right) \delta_{i j} + \partial_i \partial_j \emf .
\ee
Using these, we can define the gauge invariant (Bardeeen's) variables:
\begin{align}
    \Phi = & \phi + \frac{1}{a} \left[ a \left( \mathcal{B} - \frac{\emf'}{2} \right) \right]' , \\
    \Psi = & \psi - \frac{a'}{a} \left( \mathcal{B} - \frac{\emf'}{2} \right) + \frac{1}{6} \partial_k^2 \emf , \\
    \Theta = & \theta + \left( \mathcal{B} - \frac{\emf'}{2} \right) (\omega_0)' .
\end{align}
The lagrangians describing gauge-invariant scalar and tensor (helicity-2) perturbations in (\ref{Zwiebach_action}), on FRW background take the following shape:
\begin{widetext}
\begin{align}
	\nonumber \mathcal{L}^{EGB}_{SS} = & \frac{a^{n-2}}{2} \Big( T_{\tau \tau} + T_{xx} \Big) \Phi^2
	- (n-2) a^{n-2} \left[ (n-1) \Big( \Psi' + \mathcal{H} \Phi \Big)^2 - (n-3) \left( \partial_i \Psi \right)^2 - 2 \Phi \partial_i^2 \Psi \right] \\
	\nonumber {} & - 2 \kappa^2 (n-2) a^{n-4} \mathcal{H}^2 \left[ (n-1) \Big( \Psi' + \mathcal{H} \Phi \Big)^2 - (n-3) \left( 1 - \frac{2 \epsilon}{n-3} \right) \left( \partial_i \Psi \right)^2 - 2 \Phi \partial_i^2 \Psi \right] \\
	{} & + \frac{a^{n-2}}{2} \Big[ \left( \Theta ' \right)^2 - \left( \partial_i \Theta \right)^2 
	- a^2 \frac{d^2 V}{d \omega^2_0} \Theta^2 \Big]  + a^{n-2} \omega_0' \Theta \Big( \Phi' + (n-1) \Psi' \Big)
	- 2 a^n \frac{d V}{d \omega_0} \Theta \Phi , \label{zwei_scalar_perts} \\
    \mathcal{L}^{EGB}_{TT} = & \frac{a^{n-2}}{4} \left( 1 + \frac{2 \kappa^2}{a^2 } \mathcal{H}^2 \right) \left( h_{ij}' \right)^2 
    - \frac{a^{n-2}}{4} \left( 1 + \frac{2 \kappa^2}{a^2 } \left( 1 - \frac{2 \epsilon}{n-3} \right) \mathcal{H}^2 \right) \left( \partial_k h_{ij} \right)^2 . \label{zwei_tensor_perts}
\end{align}
\end{widetext}
The last formula was obtained using the \texttt{xPert} and \texttt{xPand} packages \cite{Brizuela:2008ra,Pitrou:2013hga}. Note that the tensor perturbation propagates (sub-)luminally $(\epsilon \equiv 1 - \mathcal{H}'/\mathcal{H}^2 \geq 0)$:
\bb
v_h^2 = \frac{1 + 2 \kappa^2 \left( 1 - \frac{2 \epsilon}{n-3} \right) \frac{\mathcal{H}^2}{a^2} }{1 + 2 \kappa^2 \frac{\mathcal{H}^2}{a^2 } } \leq 1 .
\ee
The latter is expected, as we chose the sign in front of $\kappa^2$ from the string effective action \cite{Zwiebach:1985uq,Gross:1986mw}, which admits the Wilsonian UV completion.


\vspace{.1in}
\noindent
{\it The effective action describing the trace anomaly.} -- We will focus on the following effective action describing the trace anomaly \cite{Gabadadze:2023quw}:
\begin{align}\label{Lag_beg}
	\nonumber {} & \mathcal{L} = \sqrt{g} \Bigg\{ R - \bar{M}^2 e^{-2 \sigma} R 
	- 6 \bar{M}^2 e^{-2 \sigma} \left( \partial \sigma \right)^2 \\
	\nonumber {} & - \gamma^2 \Big[ \sigma E + 4 G^{\mu \nu} \partial_\mu \sigma \partial_\nu \sigma 
	- 4 \left( \partial \sigma \right)^2 \nabla^2 \sigma + 2 \left( \partial \sigma \right)^4 \Big] \\
	{} & - \frac12 \left( \partial \omega \right)^2 - V (\omega) \Bigg\} .
\end{align}
Here $G_{\mu \nu}$ is the Einstein  tensor. The first term on line one of (\ref{Lag_beg}) is the Einstein--Hilbert term; the second and third terms are introduced to give the scalar field $\sigma$ a kinetic term. Their unconventional form is necessary to keep $\sigma$ out of the anomalous equation (\ref{cov_anom_eq}). The second line is necessary to generate the trace anomaly \cite{Riegert:1984kt,Fradkin:1983tg}. In the third line, we introduced a canonical scalar field that is supposed to source Einstein--Hilbert in the absence of the rest. The new scale $\bar{M}$ defines the cut-off of the effective theory on a flat background. For the theory defined by (\ref{Lag_beg}) to remain self-consistent, we need $\bar{M} e^{- \sigma} < 1$. The constant $\gamma^2$ (usually refferred as $a$, see eg. \cite{Hawking:2000bb}) is related to the number of degrees of freedom in the Weyl invariant sector, but in this letter we regard it as a free parameter. The full effective action of the anomaly contains the anomalyon $\times$ Weyl square term, $\sqrt{g} \sigma W_{\mu \nu \alpha \beta} W^{\mu \nu \alpha \beta} $, in addition to (\ref{Lag_beg}). We will ignore the latter as it's not relevant to our discussion. 

There exists \cite{Gabadadze:2023quw} a linear combination of the variations of (\ref{Lag_beg}) wrt to the metric and $\sigma$ that can reproduce the trace anomaly \cite{Capper:1975ig,Duff:1977ay}:
\bb\label{cov_anom_eq}
2 R - \gamma^2 E = - \left[ T \right]\,.
\ee
We call this the anomalous equation. The latter coincides with (\ref{trace_EGB}) if we replace $\kappa^2 \leftrightarrow - \gamma^2$ and take the limit $n \rightarrow 4$.

Let's again focus on the spatially flat FRW (\ref{FRW_ansatz}).
The equations of motion describing the background can be brought to the following simple form (for the details, see Appendix A in \cite{Gabadadze:2024wgj}):
\bb\label{friedmann_eq}
3 \mathcal{H}^2 - 3 \gamma^2 \frac{\mathcal{H}^4}{a^2} = \frac12 T_{\tau \tau} + \frac{\rho_c}{2 a^2}\, ,
\ee
\bb\label{sigma_H_eq}
\bar{M}^2 e^{-2 \sigma} \left( \sigma' - \mathcal{H} \right)^2  - \frac{\gamma^2}{a^2} \left( \sigma' - \mathcal{H} \right)^4
= \frac{\rho_c}{6 a^2} \,,
\ee
\bb
\left( T_\tau^\tau \right)' + 3 \mathcal{H} \Big( T_\tau^\tau - T_x^x \Big) = 0 \,.
\ee

Notice that the equation of motion of $\sigma$, (\ref{sigma_H_eq}), is of the first order because we already integrated it once. The respective integration constant, $\rho_c$, is referred to in the older literature as the Casimir energy or dark radiation \cite{Fischetti:1979ue,Starobinsky:1980te,Hawking:2000bb}. From this set of equations, we see that the only role of the anomalyon, $\sigma$, in the Friedmann equation (\ref{friedmann_eq}) is to generate a new source -- the Casimir energy. The latter also sources the difference $\sigma' - \mathcal{H}$. Setting the Casimir energy $\rho_c = 0$ erases all the trace of the anomalyon from the Friedman eq. (\ref{friedmann_eq}), while the latter takes the form of what one would expect in EGB theory (\ref{FriedmanN_EGB}) with the replacement $\kappa^2 \leftrightarrow - \gamma^2$. When $\rho_c = 0$, the eq. (\ref{sigma_H_eq}) can be trivially integrated:
\bb\label{sigma_sol}
\sigma = \sigma_0 = \log \left( \frac{\sqrt{12} \bar{M} }{c} \,a \right) .
\ee
The integration constant $c$ defines the value of the scale factor for which $\sigma$ vanishes.
Let's now study the perturbations on this solution and show that the resemblance between EGB and the effective action (\ref{Lag_beg}) extends beyond the background equations. In addition to (\ref{perts_on_FRW}) and (\ref{pert_on_omega}) we define the perturbation on (\ref{sigma_sol}) and its gauge invariant version as:
\bb
\sigma = \sigma_0 + \pi ,
\5\5
\Pi = \pi + \left( \mathcal{B} - \frac{\emf'}{2} \right) (\sigma_0)' .
\ee
The lagrangians of the scalar and tensor perturbations in terms of these gauge invariant variables:
\begin{widetext}
\begin{align}
	\nonumber \mathcal{L}_{SS} = & \frac{a^2}{2} \Big( T_{\tau \tau} + T_{xx} \Big) \Phi^2 - 2 a^2 \left[ 3 \left( \Psi' + \mathcal{H} \Phi \right)^2 - 
     \left( \partial_i \Psi \right)^2 -2 \Phi \partial_i^2 \Psi \right]  \\
	\nonumber {} & + 4 \gamma^2 \mathcal{H}^2 \left[ 3 \left( \Psi' + \mathcal{H} \Phi \right)^2  - \left( 1 - 2 \epsilon \right) \left( \partial_i \Psi \right)^2
	- 2 \Phi \partial_i^2 \Psi \right] \\
	\nonumber {} & + \frac{a^2}{2} \Big[ (\Theta ')^2 - \left( \partial_i \Theta \right)^2 - a^2 \frac{d^2 V}{d \omega^2_0} \Theta^2 \Big]
	+ a^2 \omega_0' \Theta \Big( \Phi' + 3 \Psi' \Big) - 2 a^4 \frac{d V}{d \omega_0} \Theta \Phi \\
    {} & + \frac{c^2}{12} \Big[ 6 \left( \Psi' + \Pi' \right)^2 - 6 \left( \partial_i \Pi \right)^2 
	- 2 \left( \partial_i \Psi \right)^2
	- 4 \Phi \partial_i^2 \Psi 
	- 4 \Phi \partial_i^2 \Pi + 8 \Psi \partial_i^2 \Pi \Big] , \label{scalar_perts_bardeen} \\
	\mathcal{L}_{TT} = & \frac{a^2}{4} \Big( 1 - \frac{c^2}{12 a^2} - 2 \gamma^2 \frac{\mathcal{H}^2}{a^2} \Big) \Big( h_{i j}' \Big)^2
     - \frac{a^2}{4} \Big( 1 - \frac{c^2}{12 a^2} - 2 \gamma^2 \left( 1 - 2 \epsilon \right) \frac{\mathcal{H}^2}{a^2} \Big) 
	\Big( \partial_k h_{i j} \Big)^2 . \label{tensor_perts}
\end{align}
\end{widetext}
The first three lines in lagrangian of the scalar perturbations (\ref{scalar_perts_bardeen}) coincide with scalar perturbations of the EGB (\ref{zwei_scalar_perts}) once $\kappa^2 \leftrightarrow - \gamma^2$ and $n \rightarrow 4$. Additional terms in (\ref{scalar_perts_bardeen}) on line four exist due to the mixing between anomalyon and metric scalar perturbations. Lovelock's theorem prohibits the existence of the exact EGB in 4D, and these additional terms deviating from (\ref{zwei_scalar_perts}) manifest this fact. The resemblance exists in the tensor sector as well. Formulas (\ref{tensor_perts}) and (\ref{zwei_tensor_perts}) are identical if the terms $\propto c^2$ are ignored in (\ref{tensor_perts}) and replacement and limit discussed above are done in (\ref{zwei_tensor_perts}). The deviations $\propto c^2$ exist due to the unconventional kinetic term of the anomalyon, and in an expanding universe, they may quickly become subdominant compared to the rest.

In the Supplemental Material, we compute cubic perturbations in both the scalar-scalar-scalar and tensor-scalar-scalar sectors. We demonstrate that the resemblance between the two theories extends beyond linear order to include nonlinear interactions. Specifically, we show that interactions involving three derivatives or fewer are identical in both theories. Differences between the two theories emerge at the level of interactions involving four derivatives. Unlike EGB (\ref{Zwiebach_action}), the trace anomaly action (\ref{Lag_beg}) does not contain four-derivative interactions within the purely metric sector. Instead, the four-derivative interactions of the anomalyon perturbations and their couplings with metric perturbations serve as counterparts to the purely metric four-derivative interactions present in EGB (\ref{Zwiebach_action}).


\vspace{.1in}
\noindent
{\it Summary and discussion.} -- We have shown that the newly proposed effective action (\ref{Lag_beg}) describing the trace anomaly can resemble many aspects of the Einstein--Gauss--Bonnet theory. The covariant evidence is the coincidence of the trace of the metric variation in EGB and the anomalous equation (\ref{cov_anom_eq}). Then we restricted analysis on spatially flat FRW and showed that there exists a solution for the anomalyon for which the Friedmann equations between the two theories are identical and the perturbations existing in the effective action (\ref{Lag_beg}) contain those that exist in EGB (\ref{Zwiebach_action}). The additional mixings between the perturbations that arise in the effective action (\ref{Lag_beg}), compared to (\ref{Zwiebach_action}), are the price we have to pay for the features discussed above, as exact EGB can't exist in 4D. The trace anomaly action (\ref{Lag_beg}) is an intrinsically 4D theory, so there's no need to introduce any divergent coefficient or take any limits to have the EGB-like features; they happen naturally. One very distinct aspect between the two theories is that the helicity-2 mode is sub-luminal in EGB and (seemingly) super-luminal in (\ref{Lag_beg}). Flipping the sign in front of $\gamma^2$ resolves this issue, but in that case, the problem will reappear in the anomalyon sector. In \cite{Gabadadze:2024wgj}, we argued that the seeming super-luminality of the helicity-2 mode does not create any inconsistency within the limits of the EFT. The background-dependent results of the trace anomaly effective action might be a consequence of the fact that FRW is conformally flat, and anomalyon is the Nambu-Goldstone boson of the spontaneously broken conformal symmetry \cite{Gabadadze:2020tvt}. It remains to be seen if other solutions of the anomalyon, similar to (\ref{sigma_sol}), exist such that the resemblance between the two theories extends beyond conformally flat backgrounds. This will be the subject of the future research.
In \cite{Bonifacio:2020vbk}, it was shown that equation (\ref{Lag_beg}) with $\bar{M} = 0$ arises as the $n \rightarrow 4$ limit of (\ref{Zwiebach_action}). This might suggest that the results discussed in this letter are merely trivial extensions of those presented in \cite{Bonifacio:2020vbk}. However, it is important to emphasize that the anomalyon kinetic terms $\propto \bar{M}$ are leading modifications of the EH term compared to sub-leading ones $\propto \gamma^2$, and the existence of the background solution (\ref{sigma_sol}) (which leads to the previously discussed results) relies fundamentally on these kinetic terms. Any deviation from this background destroys the correspondences established in this letter.
Last but not least, there exists a similar class of theories, sometimes referred to as Einstein-Scalar-Gauss-Bonnet (ESGB) theories, see e.g., \cite{Torii:1996yi,Ayzenberg:2014aka,Deich:2022vna,Elder:2022rak,East:2022rqi}, in which a canonical scalar field with a potential couples to the Gauss--Bonnet invariant. In contrast to anomalyon, such a scalar will necessarily source the dynamics of the space-time, so the difference between (\ref{Zwiebach_action}) and ESGB exists even at the level of background equations. 


\vspace{.1in}
\noindent
{\it Acknowledgements.} --
	I would like to thank Gregory Gabadadze, Paul Steinhardt, David Spergel, Alejandro C\'ardenas-Avenda\~no, Otari Sakhelashvili and Suvendu Giri for many useful discussions, and Jeremy Sakstein for his help with the \texttt{xPand} package. This work is supported by the Simons Foundation grant number 654561.


\bibliographystyle{utphys}
\bibliography{refs}

\providecommand{\href}[2]{#2}\begingroup\raggedright\begin{thebibliography}{10}

\bibitem{Starobinsky:1980te}
A.~A. Starobinsky, ``{A New Type of Isotropic Cosmological Models Without
  Singularity},'' \href{http://dx.doi.org/10.1016/0370-2693(80)90670-X}{{\em
  Phys. Lett. B} {\bfseries 91} (1980) 99--102}.

\bibitem{Stelle:1976gc}
K.~S. Stelle, ``{Renormalization of Higher Derivative Quantum Gravity},''
  \href{http://dx.doi.org/10.1103/PhysRevD.16.953}{{\em Phys. Rev. D}
  {\bfseries 16} (1977) 953--969}.

\bibitem{Lovelock:1971yv}
D.~Lovelock, ``{The Einstein tensor and its generalizations},''
  \href{http://dx.doi.org/10.1063/1.1665613}{{\em J. Math. Phys.} {\bfseries
  12} (1971) 498--501}.

\bibitem{Lovelock:1972vz}
D.~Lovelock, ``{The four-dimensionality of space and the einstein tensor},''
  \href{http://dx.doi.org/10.1063/1.1666069}{{\em J. Math. Phys.} {\bfseries
  13} (1972) 874--876}.

\bibitem{Glavan:2019inb}
D.~Glavan and C.~Lin, ``{Einstein-Gauss-Bonnet Gravity in Four-Dimensional
  Spacetime},'' \href{http://dx.doi.org/10.1103/PhysRevLett.124.081301}{{\em
  Phys. Rev. Lett.} {\bfseries 124} no.~8, (2020) 081301},
  \href{http://arxiv.org/abs/1905.03601}{{\ttfamily arXiv:1905.03601 [gr-qc]}}.

\bibitem{Tomozawa:2011gp}
Y.~Tomozawa, ``{Quantum corrections to gravity},''
  \href{http://arxiv.org/abs/1107.1424}{{\ttfamily arXiv:1107.1424 [gr-qc]}}.

\bibitem{Cognola:2013fva}
G.~Cognola, R.~Myrzakulov, L.~Sebastiani, and S.~Zerbini, ``{Einstein gravity
  with Gauss-Bonnet entropic corrections},''
  \href{http://dx.doi.org/10.1103/PhysRevD.88.024006}{{\em Phys. Rev. D}
  {\bfseries 88} no.~2, (2013) 024006},
  \href{http://arxiv.org/abs/1304.1878}{{\ttfamily arXiv:1304.1878 [gr-qc]}}.

\bibitem{Bonifacio:2020vbk}
J.~Bonifacio, K.~Hinterbichler, and L.~A. Johnson, ``{Amplitudes and 4D
  Gauss-Bonnet Theory},''
  \href{http://dx.doi.org/10.1103/PhysRevD.102.024029}{{\em Phys. Rev. D}
  {\bfseries 102} no.~2, (2020) 024029},
  \href{http://arxiv.org/abs/2004.10716}{{\ttfamily arXiv:2004.10716
  [hep-th]}}.

\bibitem{Gurses:2020ofy}
M.~G\"urses, T.~c. \c{S}i\c{s}man, and B.~Tekin, ``{Is there a novel
  Einstein\textendash{}Gauss\textendash{}Bonnet theory in four dimensions?},''
  \href{http://dx.doi.org/10.1140/epjc/s10052-020-8200-7}{{\em Eur. Phys. J. C}
  {\bfseries 80} no.~7, (2020) 647},
  \href{http://arxiv.org/abs/2004.03390}{{\ttfamily arXiv:2004.03390 [gr-qc]}}.

\bibitem{Gurses:2020rxb}
M.~Gurses, T.~c. \c{S}i\c{s}man, and B.~Tekin, ``{Comment on
  ''Einstein-Gauss-Bonnet Gravity in 4-Dimensional Space-Time''},''
  \href{http://dx.doi.org/10.1103/PhysRevLett.125.149001}{{\em Phys. Rev.
  Lett.} {\bfseries 125} no.~14, (2020) 149001},
  \href{http://arxiv.org/abs/2009.13508}{{\ttfamily arXiv:2009.13508 [gr-qc]}}.

\bibitem{Capper:1975ig}
D.~M. Capper and M.~J. Duff, ``{Conformal Anomalies and the Renormalizability
  Problem in Quantum Gravity},''
  \href{http://dx.doi.org/10.1016/0375-9601(75)90030-4}{{\em Phys. Lett. A}
  {\bfseries 53} (1975) 361}.

\bibitem{Duff:1977ay}
M.~J. Duff, ``{Observations on Conformal Anomalies},''
  \href{http://dx.doi.org/10.1016/0550-3213(77)90410-2}{{\em Nucl. Phys. B}
  {\bfseries 125} (1977) 334--348}.

\bibitem{Riegert:1984kt}
R.~J. Riegert, ``{A Nonlocal Action for the Trace Anomaly},''
  \href{http://dx.doi.org/10.1016/0370-2693(84)90983-3}{{\em Phys. Lett. B}
  {\bfseries 134} (1984) 56--60}.

\bibitem{Fradkin:1983tg}
E.~S. Fradkin and A.~A. Tseytlin, ``{Conformal Anomaly in Weyl Theory and
  Anomaly Free Superconformal Theories},''
  \href{http://dx.doi.org/10.1016/0370-2693(84)90668-3}{{\em Phys. Lett. B}
  {\bfseries 134} (1984) 187}.

\bibitem{Gabadadze:2023quw}
G.~Gabadadze, ``{A new gravitational action for the trace anomaly},''
  \href{http://dx.doi.org/10.1016/j.physletb.2023.138031}{{\em Phys. Lett. B}
  {\bfseries 843} (2023) 138031},
  \href{http://arxiv.org/abs/2301.13265}{{\ttfamily arXiv:2301.13265
  [hep-th]}}.

\bibitem{Fernandes:2021dsb}
P.~G.~S. Fernandes, ``{Gravity with a generalized conformal scalar field:
  theory and solutions},''
  \href{http://dx.doi.org/10.1103/PhysRevD.103.104065}{{\em Phys. Rev. D}
  {\bfseries 103} no.~10, (2021) 104065},
  \href{http://arxiv.org/abs/2105.04687}{{\ttfamily arXiv:2105.04687 [gr-qc]}}.

\bibitem{Gabadadze:2024wgj}
G.~Gabadadze, D.~N. Spergel, and G.~Tukhashvili, ``{Inflation with the trace
  anomaly action and primordial black holes},''
  \href{http://dx.doi.org/10.1103/PhysRevD.111.063529}{{\em Phys. Rev. D}
  {\bfseries 111} no.~6, (2025) 063529},
  \href{http://arxiv.org/abs/2411.16834}{{\ttfamily arXiv:2411.16834
  [hep-th]}}.

\bibitem{Misner:1973prb}
C.~W. Misner, K.~S. Thorne, and J.~A. Wheeler, {\em {Gravitation}}.
\newblock W. H. Freeman, San Francisco, 1973.

\bibitem{Zwiebach:1985uq}
B.~Zwiebach, ``{Curvature Squared Terms and String Theories},''
  \href{http://dx.doi.org/10.1016/0370-2693(85)91616-8}{{\em Phys. Lett. B}
  {\bfseries 156} (1985) 315--317}.

\bibitem{Gross:1986mw}
D.~J. Gross and J.~H. Sloan, ``{The Quartic Effective Action for the Heterotic
  String},'' \href{http://dx.doi.org/10.1016/0550-3213(87)90465-2}{{\em Nucl.
  Phys. B} {\bfseries 291} (1987) 41--89}.

\bibitem{Mukhanov:1990me}
V.~F. Mukhanov, H.~A. Feldman, and R.~H. Brandenberger, ``{Theory of
  cosmological perturbations. Part 1. Classical perturbations. Part 2. Quantum
  theory of perturbations. Part 3. Extensions},''
  \href{http://dx.doi.org/10.1016/0370-1573(92)90044-Z}{{\em Phys. Rept.}
  {\bfseries 215} (1992) 203--333}.

\bibitem{Brizuela:2008ra}
D.~Brizuela, J.~M. Martin-Garcia, and G.~A. Mena~Marugan, ``{xPert: Computer
  algebra for metric perturbation theory},''
  \href{http://dx.doi.org/10.1007/s10714-009-0773-2}{{\em Gen. Rel. Grav.}
  {\bfseries 41} (2009) 2415--2431},
  \href{http://arxiv.org/abs/0807.0824}{{\ttfamily arXiv:0807.0824 [gr-qc]}}.

\bibitem{Pitrou:2013hga}
C.~Pitrou, X.~Roy, and O.~Umeh, ``{xPand: An algorithm for perturbing
  homogeneous cosmologies},''
  \href{http://dx.doi.org/10.1088/0264-9381/30/16/165002}{{\em Class. Quant.
  Grav.} {\bfseries 30} (2013) 165002},
  \href{http://arxiv.org/abs/1302.6174}{{\ttfamily arXiv:1302.6174
  [astro-ph.CO]}}.

\bibitem{Hawking:2000bb}
S.~W. Hawking, T.~Hertog, and H.~S. Reall, ``{Trace anomaly driven
  inflation},'' \href{http://dx.doi.org/10.1103/PhysRevD.63.083504}{{\em Phys.
  Rev. D} {\bfseries 63} (2001) 083504},
  \href{http://arxiv.org/abs/hep-th/0010232}{{\ttfamily arXiv:hep-th/0010232}}.

\bibitem{Fischetti:1979ue}
M.~V. Fischetti, J.~B. Hartle, and B.~L. Hu, ``{Quantum Effects in the Early
  Universe. 1. Influence of Trace Anomalies on Homogeneous, Isotropic,
  Classical Geometries},''
  \href{http://dx.doi.org/10.1103/PhysRevD.20.1757}{{\em Phys. Rev. D}
  {\bfseries 20} (1979) 1757--1771}.

\bibitem{Gabadadze:2020tvt}
G.~Gabadadze and G.~Tukhashvili, ``{Conformal/Poincar\'e Coset, cosmology, and
  descendants of Lovelock terms},''
  \href{http://dx.doi.org/10.1103/PhysRevD.102.024054}{{\em Phys. Rev. D}
  {\bfseries 102} no.~2, (2020) 024054},
  \href{http://arxiv.org/abs/2005.01729}{{\ttfamily arXiv:2005.01729
  [hep-th]}}.

\bibitem{Torii:1996yi}
T.~Torii, H.~Yajima, and K.-i. Maeda, ``{Dilatonic black holes with
  Gauss-Bonnet term},'' \href{http://dx.doi.org/10.1103/PhysRevD.55.739}{{\em
  Phys. Rev. D} {\bfseries 55} (1997) 739--753},
  \href{http://arxiv.org/abs/gr-qc/9606034}{{\ttfamily arXiv:gr-qc/9606034}}.

\bibitem{Ayzenberg:2014aka}
D.~Ayzenberg and N.~Yunes, ``{Slowly-Rotating Black Holes in
  Einstein-Dilaton-Gauss-Bonnet Gravity: Quadratic Order in Spin Solutions},''
  \href{http://dx.doi.org/10.1103/PhysRevD.90.044066}{{\em Phys. Rev. D}
  {\bfseries 90} (2014) 044066},
  \href{http://arxiv.org/abs/1405.2133}{{\ttfamily arXiv:1405.2133 [gr-qc]}}.
  [Erratum: Phys.Rev.D 91, 069905 (2015)].

\bibitem{Deich:2022vna}
A.~Deich, A.~C\'ardenas-Avenda\~no, and N.~Yunes, ``{Chaos in quadratic
  gravity},'' \href{http://dx.doi.org/10.1103/PhysRevD.106.024040}{{\em Phys.
  Rev. D} {\bfseries 106} no.~2, (2022) 024040},
  \href{http://arxiv.org/abs/2203.00524}{{\ttfamily arXiv:2203.00524 [gr-qc]}}.

\bibitem{Elder:2022rak}
B.~Elder and J.~Sakstein, ``{Mapping the weak field limit of
  scalar-Gauss-Bonnet gravity},''
  \href{http://dx.doi.org/10.1103/PhysRevD.107.044006}{{\em Phys. Rev. D}
  {\bfseries 107} no.~4, (2023) 044006},
  \href{http://arxiv.org/abs/2210.10955}{{\ttfamily arXiv:2210.10955 [gr-qc]}}.

\bibitem{East:2022rqi}
W.~E. East and F.~Pretorius, ``{Binary neutron star mergers in
  Einstein-scalar-Gauss-Bonnet gravity},''
  \href{http://dx.doi.org/10.1103/PhysRevD.106.104055}{{\em Phys. Rev. D}
  {\bfseries 106} no.~10, (2022) 104055},
  \href{http://arxiv.org/abs/2208.09488}{{\ttfamily arXiv:2208.09488 [gr-qc]}}.

\end{thebibliography}\endgroup

\clearpage
\pagebreak
\widetext
\begin{center}
\textbf{\large Supplemental Material: On the analogue of Einstein--Gauss--Bonnet theory in 3+1 dimensions}
\end{center}

\setcounter{equation}{0}
\setcounter{figure}{0}
\setcounter{table}{0}
\makeatletter
\renewcommand{\theequation}{S\arabic{equation}}
\renewcommand{\thefigure}{S\arabic{figure}}
\renewcommand{\bibnumfmt}[1]{[S#1]}

The cubic perturbations presented below are derived in the Newtonian conformal gauge ($\mathcal{B}=\emf=0$). The equations are color-coded to facilitate the identification and comparison of their respective components. These results were obtained using the packages \texttt{xPert} and \texttt{xPand}~\cite{Brizuela:2008ra,Pitrou:2013hga}. Perturbations of the canonical scalar field $\omega$ are neglected, as they are irrelevant to our discussion.

\section*{The scalar-scalar-scalar part of the cubic perturbations}

The Lagrangian of the scalar cubic perturbations in EGB theory (\ref{Zwiebach_action}):
\begin{align}\label{cub_perts_SSS_EGB}
  \nonumber \mathcal{L}_{SSS}^{EGB} = & {\color{OliveGreen}{ \frac12 (n-2) a^{n-2} \Big\{ - \frac{1}{3} (n-5) (n-1) \Big[ - 6 \psi \left( \psi' \right)^2 + (n-3)^2 \mathcal{H}^2 \psi^3 + 2 (n-3) \mathcal{H}' \psi^3 \Big] }} \\
  \nonumber {} & ~~~~~~~~~~~~~~~~~~~~~ {\color{OliveGreen}{ + 3 (n-1)^2 \mathcal{H}^2 \psi \phi^2 + 6 (n-1) \mathcal{H} \psi' \phi^2 + 5 (n-1) \mathcal{H}^2 \phi^3 }} \\
  \nonumber {} & ~~~~~~~~~~~~~~~~~~~~~ {\color{OliveGreen}{ + (n-3) (n-1)^2 \mathcal{H}^2 \phi \psi^2 + 4 (n-3) (n-1) \mathcal{H} \phi \psi \psi' + 2 (n-1) \phi \left( \psi' \right)^2 }} \\
  \nonumber {} & ~~~~~~~~~~~~~~~~~~~~~ {\color{OliveGreen}{ - 2 (n-7) (n-3) \psi \left( \partial_i \psi \right)^2 - 2 \phi^2 \partial_i^2 \psi
- 2 (n-7) \phi \left( \partial_i \psi \right)^2 - 4 (n-5) \phi \psi \partial_i^2 \psi \Big\} }} \\
  \nonumber {} & {\color{Maroon}{ + \kappa^2 a^{n-4} \Big\{ \frac{1}{6} (n-2) (n-1) \Big[
8 \mathcal{H} \left( \psi' \right)^3 + 12 (n-5) \mathcal{H}^2 \psi \left( \psi' \right)^2 - (n-3) (n-5)^2 \mathcal{H}^4 \psi^3 }} \\
    \nonumber {} & ~~~~~~~~~~~~~~~~~~~~~~~~~~~~~~~~~~~~ {\color{Maroon}{ - 4 (n-3) (n-5) \mathcal{H}^2 \mathcal{H}' \psi^3 + 35 \mathcal{H}^4 \phi^3
    + 15 (n-1) \mathcal{H}^4 \psi \phi^2 + 60 \mathcal{H}^3 \psi' \phi^2  }} \\
  \nonumber {} &   ~~~~~~~~~~~~~~~~~~~~~~~~~~~~~~~~~~~~ {\color{Maroon}{ + 3 (n-1) (n-3) \mathcal{H}^4 \phi \psi^2 + 24 (n-3) \mathcal{H}^3 \phi \psi \psi' + 36 \mathcal{H}^2 \phi \left( \psi' \right)^2 \Big] }} \\
  \nonumber {} & ~~~~~~~~~~~~~ {\color{Maroon}{ + \frac12 (n-2) \Big[ - 4 (n-7) (n-5) \mathcal{H}^2 \psi \left( \partial_i \psi \right)^2 - 8 (n-7) \mathcal{H}' \psi \left( \partial_i \psi \right)^2 - 12 \mathcal{H}^2 \phi^2 \partial_i^2 \psi }} \\
  \nonumber {} & ~~~~~~~~~~~~~~~~~~~~~~~~~~~~~ {\color{Maroon}{ - 4 (n-7) \mathcal{H}^2 \phi \left( \partial_i \psi \right)^2  - 16 \mathcal{H} \phi \psi' \partial_i^2 \psi
- 8 (n-5) \mathcal{H}^2 \phi \psi \partial_i^2 \psi \Big] \Big\} }} \\
   {} & {\color{Cyan}{ + \kappa^2 a^{n-4} \Big\{ - 4 (n-2) \left( \psi' \right)^2 \partial_i^2 \psi 
+ 2 (n-5) \left( \partial_i \psi \right)^2 \partial_j^2 \psi + 4 \phi \Big[ \left( \partial_i^2 \psi \right)^2
- \left( \partial_i \partial_j \psi \right)^2 \Big] \Big\} }} .
\end{align}
The {\color{OliveGreen}{green terms (lines 1-4)}} in (\ref{cub_perts_SSS_EGB}) are generated from the Einstein--Hilbert part of the action. The {\color{Maroon}{purple (lines 5-9)}} and {\color{Cyan}{cyan (line 10)}} terms are generated from the Gauss--Bonnet term. The Lagrangian of the scalar cubic perturbations in the effective theory of the trace anomaly (\ref{Lag_beg}):
\begin{align}\label{cub_perts_SSS_anom}
  \nonumber \mathcal{L}_{SSS} = & {\color{OliveGreen}{ a^2 \Big\{ - 6 \psi \left( \psi' \right)^2 + \mathcal{H}^2 \psi^3 + 2 \mathcal{H}' \psi^3 
   +  27 \mathcal{H}^2 \psi \phi^2 + 18 \mathcal{H} \psi' \phi^2 + 15 \mathcal{H}^2 \phi^3 +  9 \mathcal{H}^2 \phi \psi^2 }} \\
  \nonumber {} & ~~~~~~ {\color{OliveGreen}{ + 12 \mathcal{H} \phi \psi \psi' + 6 \phi \left( \psi' \right)^2 + 6 \psi \left( \partial_i \psi \right)^2 - 2 \phi^2 \partial_i^2 \psi
+ 6 \phi \left( \partial_i \psi \right)^2 + 4 \phi \psi \partial_i^2 \psi \Big\} }} \\
  \nonumber {} & {\color{Maroon}{ - \gamma^2 \Big\{ 8 \mathcal{H} \left( \psi ' \right)^3 - 12 \mathcal{H}^2 \psi \left( \psi ' \right)^2 - \mathcal{H}^4 \psi^3 + 4 \mathcal{H}^2 \mathcal{H}' \psi^3 
  + 36 \mathcal{H}^2 \phi \left( \psi ' \right)^2 }} \\
  \nonumber {} & ~~~~~~~~ {\color{Maroon}{ + 24 \mathcal{H}^3 \phi \psi \psi' + 9 \mathcal{H}^4 \phi \psi^2
  + 60 \mathcal{H}^3 \psi' \phi^2 + 45 \mathcal{H}^4 \psi \phi^2 + 35 \mathcal{H}^4 \phi^3 }} \\
  \nonumber {} & ~~~~~~~~ {\color{Maroon}{ - 12 \mathcal{H}^2 \psi \left( \partial_i \psi \right)^2 + 24 \mathcal{H}' \psi \left( \partial_i \psi \right)^2
  + 12 \mathcal{H}^2 \phi \left( \partial_i \psi \right)^2 + 8 \mathcal{H}^2 \phi \psi \partial_i^2 \psi - 16 \mathcal{H} \phi \psi' \partial_i^2 \psi
  - 12 \mathcal{H}^2 \phi^2 \partial_i^2 \psi \Big\} }} \\
  \nonumber {} & {\color{BurntOrange}{ + \frac{c^2}{12} \Big\{ 6 \psi \left( \psi ' \right)^2 - 6 \phi \left( \psi ' \right)^2 - 6 \psi \left( \partial_i \psi \right)^2
  - 6 \phi \left( \partial_i \psi \right)^2 - 4 \phi \psi \partial_i^2 \psi + 2 \phi^2 \partial_i^2 \psi }} \\
  \nonumber {} & ~~~~~~~~ {\color{BurntOrange}{ - 12 \pi \left( \pi ' \right)^2 + 12 \pi^2 \psi'' - 18 \psi \left( \pi ' \right)^2 - 6 \phi \left( \pi ' \right)^2
  + 12 \psi \psi''  \pi - 12 \phi \psi' \pi' + 12 \pi \left( \partial_i \pi \right)^2 + 6 \psi \left( \partial_i \pi \right)^2 }} \\
  \nonumber {} & ~~~~~~~~ {\color{BurntOrange}{ - 8 \pi^2 \partial_i^2 \psi + 2 \phi \left( \partial_i \pi \right)^2 + 8 \phi \pi \partial_i^2 \pi + 12 \pi \left( \partial_i \psi \right)^2 + 8 \pi \psi \partial_i^2 \psi
  + 2 \phi^2 \partial_i^2 \pi
  - 4 \psi \partial_i \phi \partial_i \pi + 8 \phi \pi \partial_i^2 \psi \Big\} }} \\
  \nonumber {} & {\color{blue}{ - \gamma^2 \Big\{ 8 \left( \pi' \right)^2 \partial_i^2 \pi - 4 \left( \partial_i \pi \right)^2 \partial_j^2 \pi
  + 16 \psi \Big[ \left( \partial_i \pi' \right)^2 - \pi'' \partial_i^2 \pi \Big]
  + 16 \pi \Big[ \left( \partial_i \psi' \right)^2 - \psi'' \partial_i^2 \psi \Big] }} \\
  {} & ~~~~~~~~ {\color{blue}{ + 4 \psi \Big[ \left( \partial_i^2 \pi \right)^2 - \left( \partial_i \partial_j \pi \right)^2 \Big]
  - 4 \phi \Big[ \left( \partial_i^2 \pi \right)^2 - \left( \partial_i \partial_j \pi \right)^2 \Big] 
  - 8 \phi \Big[ \partial_i^2 \psi \partial_j^2 \pi - \partial_i \partial_j \psi \partial_i \partial_j \pi \Big] \Big\} }} .
\end{align}
The {\color{OliveGreen}{green (lines 1-2)}} and {\color{Maroon}{purple (lines 3-5)}} terms in (\ref{cub_perts_SSS_anom}) are identical to the {\color{OliveGreen}{green}} and {\color{Maroon}{purple}} terms found in (\ref{cub_perts_SSS_EGB}) once the replacement $\kappa^2 \leftrightarrow -\gamma^2$ is done and $n \rightarrow 4$ limit is taken. The {\color{BurntOrange}{orange terms (lines 6-8)}} in (\ref{cub_perts_SSS_anom}) exist due to the conformal kinetic mixing between the anomlayon and Einstein--Hilbert. The {\color{blue}{blue terms}} -- perturbations of the anomalyon and their mixings with the scalar perturbations of the metric (lines 9-10 in (\ref{cub_perts_SSS_anom})) are generated from the second line of (\ref{Lag_beg}). The {\color{BurntOrange}{orange}} and {\color{blue}{blue}} terms (lines 6-10 in (\ref{cub_perts_SSS_anom})) behave as if they lived on a flat background, which is a feature of the solution (\ref{sigma_sol}).

The Lagrangian (\ref{cub_perts_SSS_anom}) does not contain the four derivative, {\color{Cyan}{cyan terms (line 10)}} that exist in (\ref{cub_perts_SSS_EGB}), this is because $\sqrt{g} E $ is a total derivative in 4D and at least one of the derivatives needs to be spent by acting it on $\sigma$, as a consequence there are no four derivative interactions in the purely metric sector of the scalar perturbations in (\ref{cub_perts_SSS_anom}), but there are three and fewer derivative interactions that coincide with those existing in (\ref{cub_perts_SSS_EGB}). The four-derivative, {\color{blue}{blue interactions (lines 9-10)}} in (\ref{cub_perts_SSS_anom}) exist due the anomalyon fluctuations. It's interesting to note that if $\pi \rightarrow - \psi$ replacement is done, the {\color{blue}{blue terms (lines 9-10)}} of (\ref{cub_perts_SSS_anom}) will coincide with the {\color{Cyan}{cyan terms (line 10)}} of (\ref{cub_perts_SSS_EGB}). 


\section*{The tensor-scalar-scalar part of the cubic perturbations}

The Lagrangian of the tensor-scalar-scalar perturbations in EGB theory (\ref{Zwiebach_action}):
\begin{align}\label{cub_perts_TSS_EGB}
	\nonumber \mathcal{L}_{TSS}^{EGB} = & {\color{OliveGreen}{ (n-3) a^{n-2} \Big\{ 2 h_{ij} \partial_i \phi \partial_j \psi - (n-4) h_{ij} \partial_i \psi \partial_j \psi \Big\} }} \\
	\nonumber {} & {\color{Maroon}{ + \kappa^2 a^{n-4} \Big\{ 4 \mathcal{H} h_{ij}' \partial_i \phi \partial_j \psi - 4 (n-3) \mathcal{H}^2 h_{ij} \phi \partial_i \partial_j \psi 
	+ 2 (n-4) \Big( \mathcal{H} h_{ij}' \partial_i \psi \partial_j \psi
	+ \mathcal{H}^2 h_{ij} \partial_i \psi \partial_j \psi - \mathcal{H}' h_{ij} \partial_i \psi \partial_j \psi \Big) \Big\} }} \\
	\nonumber {} & {\color{Cyan}{ + \kappa^2 a^{n-4} \Big\{ 4 h_{ij} \Big( \psi'' \partial_i \partial_j \psi - \partial_i \psi' \partial_j \psi' \Big)
	+ \frac{4 (n-5)}{n-3} h_{ij} \Big( \partial_i \partial_k \psi \partial_j \partial_k \psi - \partial_i \partial_j \psi \partial_k^2 \psi \Big) }} \\
	{} & ~~~~~~~~~~~~~~ {\color{Cyan}{ + \frac{4}{n-3} h_{ij} \Big( \partial_i \partial_j \phi \partial_k^2 \psi + \partial_k^2 \phi \partial_i \partial_j \psi
	- 2 \partial_i \partial_k \phi \partial_j \partial_k \psi \Big) \Big\} }} .
\end{align}
Similar to the previous section, the {\color{OliveGreen}{green terms (line 1)}} in (\ref{cub_perts_TSS_EGB}) are generated from the Einstein--Hilbert part of the action, while the {\color{Maroon}{purple (line 2)}} and {\color{Cyan}{cyan (lines 3-4)}} terms are generated from the Gauss--Bonnet term. The Lagrangian of the tensor-scalar-scalar perturbations in the effective theory of the trace anomaly (\ref{Lag_beg}):
\begin{align}\label{cub_perts_TSS_anom}
	\nonumber \mathcal{L}_{TSS} = & {\color{OliveGreen}{ 2 a^2 h_{ij} \partial_i \phi \partial_j \psi }} {\color{Maroon}{ - \gamma^2 \Big\{ 4 \mathcal{H} h_{ij}' \partial_i \phi \partial_j \psi - 4 \mathcal{H}^2 h_{ij} \phi \partial_i \partial_j \psi \Big\} }} \\
	\nonumber {} & {\color{BurntOrange}{ + \frac{c^2}{12} \Big\{ - 2 h_{ij} \partial_i \phi \partial_j \psi
	- 4 h_{ij} \partial_i \phi \partial_j \pi + 6 h_{ij} \partial_i \pi \partial_j \pi + 4 h_{ij} \partial_i \psi \partial_j \pi \Big\} }} \\
	\nonumber {} & {\color{blue}{ - \gamma^2 \Big\{ - 4 h_{ij} \Big( \pi'' \partial_i \partial_j \psi + \psi'' \partial_i \partial_j \pi 
	- 2 \partial_i \pi' \partial_j \psi' \Big) - 4 h_{ij} \Big( \pi'' \partial_i \partial_j \pi - \partial_i \pi' \partial_j \pi' \Big) }} \\
	{} & ~~~~~~~~ {\color{blue}{ - 4 h_{ij} \Big( \partial_i \partial_k \pi \partial_j \partial_k \pi
	- \partial_k^2 \pi \partial_i \partial_j \pi \Big) - 4 h_{ij} \Big( \partial_k^2 \pi \partial_i \partial_j \phi + \partial_k^2 \phi \partial_i \partial_j \pi
	- 2 \partial_i \partial_k \pi \partial_j \partial_k \phi \Big) \Big\} }} .
\end{align}
The {\color{OliveGreen}{green (the first term on line 1)}} and {\color{Maroon}{purple (terms $\propto \gamma^2$ on line 1)}} interactions in (\ref{cub_perts_TSS_anom}) are identical to the {\color{OliveGreen}{green}} and {\color{Maroon}{purple}} terms found in (\ref{cub_perts_TSS_EGB}) once the replacement $\kappa^2 \leftrightarrow -\gamma^2$ is done and $n \rightarrow 4$ limit is taken. The {\color{BurntOrange}{orange terms (line 2)}} in (\ref{cub_perts_TSS_anom}) exist due to the conformal kinetic mixing between the anomlayon and Einstein--Hilbert. The {\color{blue}{blue terms}} -- perturbations of the anomalyon and their mixings with the scalar perturbations of the metric (lines 3-4 in (\ref{cub_perts_TSS_anom})) are generated from the second line of (\ref{Lag_beg}). Similar to (\ref{cub_perts_SSS_anom}), interactions that involve the fluctuation of the anomalyon behave as if the space-time was flat.

Due to the reasons explained by the end of the previous chapter, the four-derivative interactions involving purely metric perturbations are absent in (\ref{cub_perts_TSS_anom}). Instead we get the {\color{blue}{blue interactions}}, involving the anomalyon fluctuation, which reduce to {\color{Cyan}{cyan interactions}} of (\ref{cub_perts_TSS_EGB}) if $\pi \rightarrow - \psi$ replacement is done. 

The observation that {\color{blue}{blue interactions}} transition to {\color{Cyan}{cyan}} if $\pi \rightarrow - \psi$, gives rise to the speculation that the four-derivative anomalyon interactions, along with their mixings with the scalar components of the metric perturbations, serve as the counterparts to the four-derivative interactions present in EGB. This is a consequence of the fact that the anomalyon is the Nambu--Goldstone boson of spontaneously broken conformal invariance \cite{Gabadadze:2020tvt}, and $\psi$ is related to the conformal mode of the metric.

There exists a non-linear version of the replacement discussed above:
\bb
\pi \rightarrow - \psi - \frac12 \frac{\partial_i \partial_j}{\partial_k^2} \Big[ \psi \Big( h_{ij} + 2 \delta_{ij} \psi \Big) \Big] .
\ee
If this replacement is made, the orange terms in (\ref{cub_perts_SSS_anom}) and (\ref{cub_perts_TSS_anom}) disappear, same happens to the last line of (\ref{scalar_perts_bardeen}); {\color{blue}{blue interactions}} transition to {\color{Cyan}{cyan}} and we get full equivalence between the two theories (terms $\propto c^2$ remain in (\ref{tensor_perts}), but in an expanding Universe they quickly become sub-dominant).


\end{document}